\documentclass[conference]{IEEEtran}

\usepackage{pgfplotstable}

\usepackage{tikz,pgf,pgfplots}
\usetikzlibrary{calc,shadows}
\usetikzlibrary{pgfplots.groupplots}

\usepackage{times}
\usepackage{xr}

\usepackage{amsthm}
\usepackage{amsmath}    
\usepackage{amssymb}    
\usepackage{mathrsfs}   
\usepackage{epsfig}     
\usepackage{graphicx}
\usepackage{url}

\usepackage{url}
\usepackage{color}
\usepackage{booktabs}

\usepackage{subfigure}

\usepackage{amsfonts}
\usepackage{dsfont}
\usepackage{url}
\usepackage{textcomp}
\usepackage{multirow}
\usepackage[numbers,sort&compress]{natbib}

\usepackage{enumitem}

\usepackage{colortbl}

\graphicspath{{}{figs/}{../}{../figs/}{../Monograph/monograph/nowformat/}}

\newif\ifmonograph
\monographfalse

\renewcommand{\caption}[1]{\singlespacing\hangcaption{#1}\normalspacing}

\usepackage[numbers,sort&compress]{natbib}

\usepackage{amsmath}
\usepackage{amsfonts}
\usepackage{dsfont}
\usepackage{mathrsfs}

\usepackage{algorithmic}
\usepackage{algorithm}
\usepackage{multirow}

\newcommand{\ep}{\epsilon}

\newcommand{\E}{{\mathcal E}}

\newcommand{\M}{{\mathcal M}}
\newcommand{\N}{{\mathcal N}}

\newcommand{\defi}{\triangleq}

\newcommand{\one}{\mathds 1}

\newcommand{\st}{:}

\newcounter{constcount}
\newcounter{numcount}
\newcounter{thmcnt}
  \let\Oldsection\section
  
\renewcommand{\section}{\stepcounter{thmcnt}\Oldsection}

\allowdisplaybreaks

\newtheorem{theorem}{Theorem} 
\newtheorem{lemma}{Lemma} 

\newtheorem{remark}{Remark}[section] 
\newtheorem{question}{Question} 

\newcounter{examplecounter}
\newcommand{\aln}[1]{\begin{align*}#1\end{align*}}

\newcommand{\al}[1]{\begin{align}#1\end{align}}

\def\Item$#1${\item $\displaystyle#1$
   \hfill\refstepcounter{equation}(\theequation)}

\newcommand{\bea}{\begin{eqnarray}}
\newcommand{\eea}{\end{eqnarray}}
\newcommand{\beas}{\begin{eqnarray*}}
\newcommand{\eeas}{\end{eqnarray*}}

\usepackage{ifthen}
\usepackage{mathtools}

\newcommand\Tex{}
\newcommand\PR[2][\Tex]{
\ifthenelse{\equal{#1}{}}{{\mathrm{Pr}}\left(#2\right)}{\ensuremath{{\mathrm{Pr}}_{#1}\left[ #2\right]}}}

\newcommand\EX[2][\Tex]{
\ifthenelse{\equal{#1}{}}{{\mathbb E}\left[#2\right]}{\ensuremath{{\mathbb E}_{#1}\left[ #2\right]}}}

\newcommand\Var[2][\Tex]{
\ifthenelse{\equal{#1}{}}{{\mathrm{Var}}\left[#2\right]}{\ensuremath{{\mathrm{Var}}_{#1}\left[ #2\right]}}}

\renewcommand\M{M} 
\renewcommand\N{N} 
\newcommand\len{L} 

\newcommand\Ex{\mathbf E} 

\newcommand\ML{ML} 

\newcommand\Cbsc{C_{\text{BSC}}} 

\newcommand\Rbsc{R_{\text{BSC}}} 
\newcommand\Rind{R_{\text{index}}}

\newcommand\cNM{c} 

\begin{document}

\title{Capacity Results for the Noisy Shuffling Channel}

\author{  
  \IEEEauthorblockN{Ilan Shomorony}  
   \IEEEauthorblockA{University of Illinois at Urbana-Champaign
 \\ 
ilans@illinois.edu  
\vspace{-3mm}
   }
  \and 
    \IEEEauthorblockN{Reinhard Heckel}
   \IEEEauthorblockA{Rice University \\ rh43@rice.edu  \vspace{-3mm}
   }
 }


\maketitle

\begin{abstract}
Motivated by DNA-based storage, we study the \emph{noisy shuffling channel}, which can be seen as the concatenation of a standard noisy channel (such as the BSC) and a shuffling channel, which breaks the data block into small pieces and shuffles them.
This channel models a DNA storage system, by capturing two of its key aspects:
(1) the data is written onto many short DNA molecules that are stored in an unordered way and (2) the molecules are corrupted by noise at synthesis, sequencing, and during storage.
For the BSC-shuffling channel we characterize the capacity exactly (for a large set of parameters), and show that a simple index-based coding scheme is optimal.
\end{abstract}

\section{Introduction}


%
Due to its longevity and enormous information density, and thanks to rapid advances in technologies for writing (synthesis) and reading (sequencing), DNA is on track to become an attractive medium for archival data storage. 
Computer scientists and engineers have dreamed of harnessing DNA's storage capabilities already in the 60s~\cite{neiman_fundamental_1964,baum_building_1995}, and in recent years this idea has developed into an active field of research:
In 2012 and 2013 groups lead by Church~\cite{church_next-generation_2012} and Goldman~\cite{goldman_towards_2013} independently stored about a megabyte of data in DNA. 
In 2015 Grass et al.~\cite{grass_robust_2015} demonstrated that millenia long storage times are possible by protecting the data both physically and information-theoretically, and designed a robust DNA data storage scheme using modern error correcting codes. Later, in the same year, Yazdi et al~\cite{yazdi_rewritable_2015} showed how to selectively access files, and in 2017, Erlich and Zielinski~\cite{erlich_dna_2016} demonstrated that practical DNA storage can achieve very high information densities. In 2018, Organick et al.~\cite{organick_scaling_2017} scaled up these techniques and stored about 200 megabytes of data.

DNA is a long molecule made up of four nucleotides (Adenine, Cytosine, Guanine, and Thymine) and, for storage purposes, can be viewed as a string over a four-letter alphabet. 
However, due to technological constraints, it is difficult and inefficient to synthesize long strands of DNA.
Thus, data is stored on short DNA molecules which are preserved in a pool of DNA and cannot be spatially ordered.
Indeed, all the aforementioned works did store information on molecules of no longer than a few hundred nucleotides. 

Given these constraints, the simplest mathematical model for a DNA storage system is a \emph{shuffling channel}: data is encoded into $\M$ DNA molecules (i.e., strings), each of length $\len$, and the output of the channel consists of a random shuffle of the $\M$ sequences.

In our previous work \cite{DNAStorageISIT}, we studied the capacity of such a shuffling channel with one additional feature: random sampling of the shuffled sequences.
In a DNA storage system, the information is accessed via shotgun sequencing, which in essence corresponds to randomly sampling and reading molecules from the DNA pool. 
In order to formally study the capacity of this system, we considered an asymptotic regime in which $\M$ is the number of sequences/molecules stored, $\N = c \M$ sequences are sampled, the length of each sequence is small (specifically $\len = \beta \log \M$ for a fixed $\beta > 0$, and throughout $\log$ is with respect to base $2$), and $\M \to \infty$.
For this setting, and assuming an alphabet of size two for simplicity, the storage capacity, defined as the maximum number of bits that can be reliably stored per symbol (the total number of symbols is $\ML$) was shown to be
\al{
(1 - e^{-c}) (1 - 1/\beta),
\label{eq:prevcap}
}
provided that $\beta > 1$. 
If $\beta < 1$, no positive rate is achievable. 

Surprisingly, the capacity of this shuffle-and-sampling channel is achieved by a simple scheme where each sequence is given a unique index (or header), and an erasure outer code is used to protect against missing molecules in the sampling step. 
The factor $1 - e^{-\cNM}$ can be  understood as the loss due to unseen molecules, and the factor $1-1/\beta$ corresponds to the loss due to shuffling or the lack of ordering in the molecules.

In practice, however, the DNA sequences are additionally corrupted by substitutions, deletions, and insertion errors (which occur both at synthesis, sequencing, and during the storage period). 
For that reason, several recent works have studied the design of error-correcting codes tailored to the error profiles of DNA storage systems
\cite{kiah_codes_2016,yazdi_rewritable_2015,gabrys_asymmetric_2015,sala_insertions_2016,kovavcevic2018asymptotically}.
However, generalizing the capacity expression in (\ref{eq:prevcap}) to a setting in which, in addition to random shuffling and random sampling, the molecules are corrupted by noise is quite nontrivial.


\begin{figure}[b] 
	\center
       \includegraphics[width=0.95\linewidth]{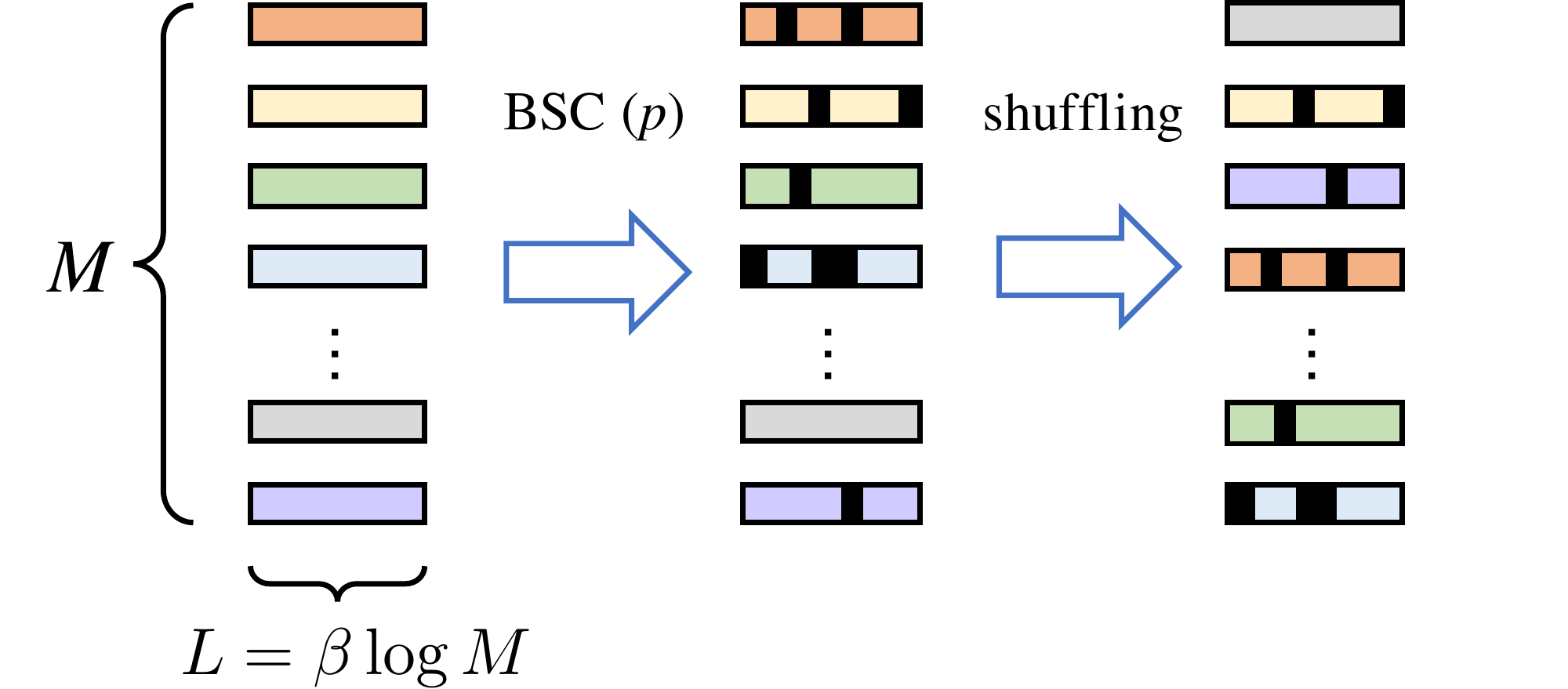} 
       \caption{Noisy (BSC) shuffling channel.\label{fig:channel}}
\end{figure}

In this work we make progress on this front by studying the capacity of a noisy shuffling channel.
We drop the random sampling constraint considered in \cite{DNAStorageISIT}, and focus on a setting in which the $\M$ sequences are first corrupted by noise and then randomly shuffled.
Since in all recently proposed systems~\cite{church_next-generation_2012,goldman_towards_2013,grass_robust_2015,yazdi_rewritable_2015,erlich_dna_2016,organick_scaling_2017}, substitution errors are more frequent than insertion and deletion errors~\cite{heckel_characterization_2018}, we focus on substitution noise.

%
%
%


{\bf Contributions:} 
We study the capacity of the noisy shuffling channel illustrated in Figure~\ref{fig:channel},
where the input/ouput sequences are binary, and the noisy channel is a Binary Symmetric Channel (BSC) with crossover probability $p$.
Our main result establishes that, for a large set of the parameters $\beta$ and $p$, 
the capacity of the BSC shuffling channel is 
\al{
C = 1 - H(p) - 1/\beta. 
\label{eq:curcap}
}
Since the capacity of a BSC channel is $\Cbsc = 1-H(p)$, this results implies that the shuffling reduces the BSC capacity by $1/\beta$ bits.


The capacity-achieving scheme consists of encoding each length-$\len$ string as a length-$\len$ codeword from a BSC capacity-achiving code, and using $\log \M$ out of its $\len \Cbsc$ information bits (prior to encoding) to store an index.
Hence, at least asymptotically, there is no rate gain in coding across the $\M$ strings, 
and index-based approaches, which were used in several practical implementations \cite{grass_robust_2015,yazdi_rewritable_2015,erlich_dna_2016,organick_scaling_2017} are optimal.


{\bf Outline: }
The paper is organized as follows.
After formalizing the problem setting in Section~\ref{sec:problem}, we state our main result and describe the achievability in Section~\ref{sec:cap}.
Section~\ref{sec:converse} is dedicated to the converse.
We conclude by discussing extensions and future directions in Section~\ref{sec:discussion}.




{\bf Related literature: }
Motivated by DNA-based storage, a few recent works have considered the problem of coding across an unordered set of strings \cite{kovavcevic2018codes,song2018sequence,LenzAnchor,lenz2018coding}.
The setting studied in all these works bears similarities with the one in this paper, but they focus on providing explicit code constructions, as opposed to characterizing the channel capacity, as we do here.

\section{Problem Setting}
\label{sec:problem}


The BSC shuffling channel, illustrated in Figure~\ref{fig:channel}, can be seen as the concatenation of 
a BSC with crossover probability $p$ and a shuffling channel, which randomly reorders the binary strings.
Formally, we will view the input to the channel as a length-$\ML$ binary string
\aln{
X^{\ML} = \left[ X_1^\len, X_2^\len, \ldots , X_\M^\len \right]
}
 (i.e., $\M$ strings of length $\len$ concatenated to form a single string of length $\M \len$).
 Similarly, 
\aln{
Y^{\ML} = \left[ Y_1^\len, Y_2^\len, \ldots, Y_\M^\len \right]
}
is the corresponding output.
We let $Z^{\ML}= \left[ Z_1^\len, \ldots, Z_\M^\len \right]$ be the random binary error pattern created by the BSC on $X^{\ML}$ and $S^\M \in \{1,\ldots,\M\}^\M$ be a vector encoding the random shuffle induced by the channel, so that 
\aln{
Y_k^\len &= X^{\len}_{S(k)} \oplus Z^{\len}_{S(k)},
}
for $k=1,\ldots,\M$, where $\oplus$ indicates elementwise modulo $2$ addition.
Similar to our previous work~\cite{DNAStorageISIT}, we let $\len = \beta \log \M$, and consider the asymptotic regime $\M \to \infty$.
As shown in~\cite{DNAStorageISIT}, this is the proper regime for positive data rates to be achievable.

A rate $R$ is said to be achievable if we have a sequence of codes, each with $2^{\M \len R}$ codewords, and whose error probability goes to zero as $\M \to \infty$.
The shuffling channel capacity $C$ is the supremum over all achievable rates.

\section{Main Result}
\label{sec:cap}

\pgfplotsset{compat=1.10}
\usepgfplotslibrary{fillbetween}

Our main result establishes that, provided that $\beta$ is large enough, treating each length-$\len$ sequence as the input to a separate BSC and encoding a unique index into each sequence is capacity optimal.
More precisely, consider code for a BSC with codewords of length $\len$ and rate $R_{\text{BSC}} = 1-H(p) - \ep$, for some small $\ep$.
Using this code, we can store $\len \Rbsc$ information bits in each length-$\len$ string.
We utilize the first $\log \M$ of those bits to encode a distinct index for each of the $\M$ strings.
Hence, in each string we can effectively store $\len \Rbsc - \log M$ data bits, and the rate of this scheme is
\al{
\frac{\M \left( \len \Rbsc - \log \M\right)}{\M \len} = \Rbsc - 1/\beta. \label{eq:schemerate}
}
Since $\ep > 0$ can be chosen arbitrarily small,
this scheme can have a rate arbitrarily close to 
\al{
\Rind = 1 - H(p) - 1/\beta, \label{eq:lower}
}
if $1 - H(p) - 1/\beta > 0$.


\vspace{1mm}

Strictly speaking, the simple index-based scheme described above needs to be modified using an outer code to guarantee that the rate in (\ref{eq:schemerate}) is actually achieved.
That is because 
there is a vanishing, but positive, probability that a given string is decoded in error, and since $M \to \infty$, the probability that at least one of the strings is decoded in error may not go to zero.
When an error occurs, the index is also read in error, which may cause a ``collision'' with a properly decoded string.
Nevertheless, since the error probability for each given string vanishes as $M \to \infty$, 
this can be fixed with an outer code across the data symbols of all the $M$ strings, and it is straightforward to see that one can achieve arbitrarily close to the rate in (\ref{eq:schemerate}).
Hence, (\ref{eq:lower}) is a lower bound to the capacity $C$.

On the other hand, a simple upper bound is given by
\al{
C \leq \min\left[ 1- H(p), 1-1/\beta \right], \label{eq:upper}
}
since both the capacity of the BSC, $\Cbsc = 1-H(p)$, and the capacity of the shuffling channel, $1-1/\beta$ \cite{DNAStorageISIT}, must be upper bounds to $C$.

Our main result improves on the upper bound in (\ref{eq:upper}), and establishes that for parameters $(p,\beta)$ in a certain regime, the lower bound in equation~\eqref{eq:lower} is actually the capacity.

\begin{theorem}
\label{thm:mainres}
If $p \leq 0.1$ and $\beta \geq 6.4$, 
\al{
C = 1 - H(p) - 1/\beta. \label{eq:capacity}
}
Moreover, if $\beta \leq 1$, $C = 0$.
\end{theorem}

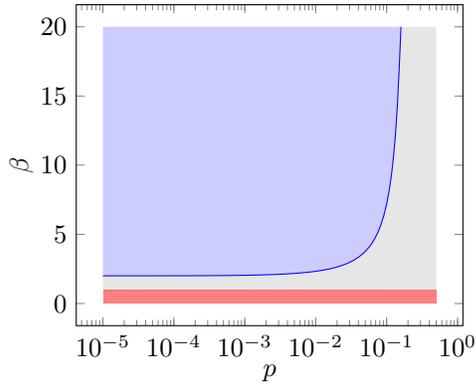
\begin{figure}[h]
\centering 
\vspace{1cm}
\begin{tikzpicture}[scale=0.75]

    \begin{semilogxaxis}[enlargelimits=0.08,xlabel=$p$,ylabel=$\beta$]
\addplot[name path=f,color=blue,mark=none,draw=blue!] table[x index=0,y index=1]{./curve.dat};   
   
   \path[name path=axis] (axis cs:0.00001,20) -- (axis cs:0.1,20);
   \addplot[blue!20] fill between[of=f and axis];

   \path[name path=axis2] (axis cs:0.00001,0.9) -- (axis cs:0.5,1) -- (axis cs:0.5,20) -- (axis cs:0.1,20);
   \addplot[gray!20] fill between[of=f and axis2];    
   
   \addplot[name path=f, fill=red, fill opacity=0.5, draw=none, mark=none]
coordinates {
    (0.00001, 0)
    (0.00001, 1)
    (0.5, 1)
    (0.5, 0)
};

\end{semilogxaxis}
\end{tikzpicture}  
\vspace{2mm}
\caption{\label{fig:capacity}
Parameter regions for which we characterized the capacity.
The capacity in the blue region is given by $C = 1 - H(p) - 1/\beta$, and the capacity in the red region (i.e., for $\beta < 1$) is $0$.
In the gray region, it is still unknown.}
\vspace{-2mm}
\end{figure}

As we will see, the converse argument presented in the next section and the capacity expression in (\ref{eq:capacity}) hold for a set of parameters $(p,\beta)$ larger than those specified in Theorem~\ref{thm:mainres}.
In fact, for all $(p,\beta)$ in the blue region of Figure~\ref{fig:capacity}, the capacity is given by (\ref{eq:capacity}).
Hence, for $p < 0.01$ for instance, we need $\beta \geq 2.35$ for (\ref{eq:capacity}) to hold.

%
%

\section{Converse}
\label{sec:converse}

Consider a sequence of codes for the noisy shuffling channel with rate $R$ and vanishing error probability.
Let 
\aln{
X^{\ML} = \left[ X_1^\len, X_2^\len, \ldots, X_\M^\len \right]
}
be the input to the channel when we choose one of the $2^{\ML R}$ codewords from one such code uniformly at random, and 
\aln{
Y^{\ML} = \left[ Y_1^\len, Y_2^\len, \ldots, Y_\M^\len \right]
}
the corresponding output.
We first observe that 
\aln{
MLR & = H\left(X^{\ML}\right) = I\left(X^{\ML};Y^{\ML}\right) - \ML \epsilon_{\M},
}
where $\epsilon_{\M} \to 0$ as $\M \to \infty$ by Fano's inequality.
Then,
\al{
ML(R-\epsilon_{\M}) & = H\left(Y^{\ML}\right) -  H\left(Y^{\ML}| X^{\ML} \right) \nonumber \\
& = H\left(Y^{\ML}\right) -  H\left(S^M,Z^{\ML},Y^{\ML} | X^{\ML}\right) \nonumber \\ 
& \quad + H\left(S^M,Z^{\ML}|X^{\ML},Y^{\ML}\right) \nonumber \\ 
& = H\left(Y^{\ML}\right) -  H\left(S^M,Z^{\ML},Y^{\ML} | X^{\ML}\right) \nonumber \\ 
& \quad + H\left(S^M|X^{\ML},Y^{\ML}\right). \label{eq:bound1}
}
The last equality follows by noticing that, given $(S^\M,X^{\ML},Y^{\ML})$, one can compute
\aln{
Z^{\len}_{S(k)} = Y_k^\len \oplus X^{\len}_{S(k)} 
}
for $k = 1,\ldots,\M$, and thus  $H\left(Z^{\ML}|X^{\ML},Y^{\ML},S^M\right) = 0$.
The second term in (\ref{eq:bound1}) can be computed as
\al{
& H\left(S^M,Z^{\ML},Y^{\ML} | X^{\ML}\right) \nonumber\\
& \quad  = H\left(S^M\right) + H\left(Z^{\ML}\right) + H\left(Y^{\ML} | X^{\ML}, S^M,Z^{\ML}\right) \nonumber\\
& \quad = \log \M! + \ML H(p) + 0 \nonumber \\
& \quad = \M \log \M + \ML H(p) + o(\ML).  \label{eq:bound2}
}


In order to finish the converse, we need to jointly bound the first and third terms in equation~(\ref{eq:bound1}).
This is the most challenging part of the proof and is summarized in the following lemma:
\begin{lemma} \label{lem1}
If $\beta$ and $p$ satisfy
\al{
1-H(2p)- \frac{2}{\beta} > 0 \label{eq:condition}
}
then it holds that 
\aln{
H\left(Y^{\ML}\right) + H\left(S^M|X^{\ML},Y^{\ML}\right) \leq \ML + o(\ML).
}
\end{lemma}
The parameter regime $(p,\beta)$ for which the inequality~\eqref{eq:condition} holds is the regime in which our capacity expression holds. 
The regime is depicted in blue in Figure~\ref{fig:capacity}. 

Combining (\ref{eq:bound1}), (\ref{eq:bound2}) and Lemma \ref{lem1}, we obtain 
\aln{
ML(R-\epsilon_{\M}) \leq \ML  - \ML H(p) - M \log \M +  o(\ML).
}
Dividing by $\ML$ and letting $\M \to \infty$ yields the converse.

\subsection{Intuition for Lemma~\ref{lem1}}

\begin{figure}[ht] 
	\center
       \includegraphics[width=0.9\linewidth]{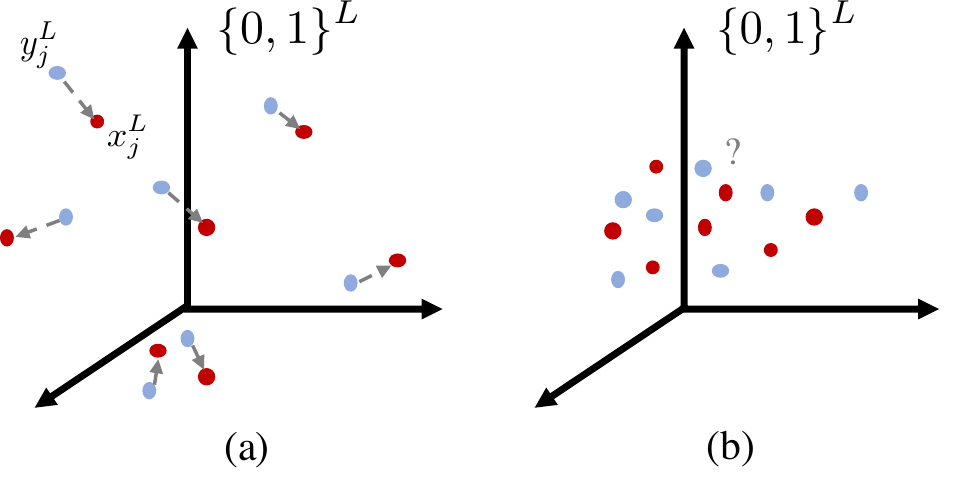} 
              \vspace{-3mm}
       \caption{Two opposite scenarios for estimating $S^M$ from $\left(X^{\ML},Y^{\ML}\right)$.\label{fig:points}}
\end{figure}

If we trivially bound each entropy term separately, we obtain 
\aln{
H\left(Y^{\ML}\right) + H\left(S^M|X^{\ML},Y^{\ML}\right) \leq \ML + \M \log \M. 
}
However, intuitively, the bound $H\left(S^M|X^{\ML},Y^{\ML}\right) \leq \M \log \M$ is too loose. 
Given $X^{\ML} = x^{\ML}$ and $Y^{\ML} = y^{\ML}$, one could estimate $S^M$ by finding for each $y_i^\len$ the $x_j^\len$ that is closest to it and guessing that $S(i) = j$.
This will be a good guess if there is no other $x_k^\len$ close to $x_j^\len$.
Hence, we can picture two opposite scenarios, illustrated in Figure~\ref{fig:points}.
In the first case, the strings $x_1^\len,\ldots,x_\M^\len$ are all distant from each other (in the Hamming sense).
Hence, the maximum likelihood estimate of $S^M$ given $X^{\ML} = x^{\ML}$ and $Y^{\ML} = y^{\ML}$ will be ``close'' to the truth and we would expect 
$H\left(S^M|X^{\ML} = x^{\ML},Y^{\ML} = y^{\ML}\right)$ to be small.
In the second case, illustrated in Fig.~\ref{fig:points}(b), several of $x_1^\len,...,x_\M^\len$ are close to each other.
So we have less information about $S^M$, and $H\left(S^M|X^{\ML} = x^{\ML},Y^{\ML} = y^{\ML}\right)$ may be large.

The term $H\left(Y^{\ML}\right)$ is maximized by making the $\left\{X_i^\len\right\}$ independent and taking values uniformly in $\{0,1\}^\len$.
Hence, in order for $H\left(Y^{\ML}\right)$ to be large, we would expect the scenario in Fig.~\ref{fig:points}(a) to be common and the scenario in Fig.~\ref{fig:points}(b) to be rare.
This creates a tension between the terms $H\left(Y^{\ML}\right)$ and $H\left(S^M|X^{\ML},Y^{\ML}\right)$, which we exploit to prove Lemma~\ref{lem1}.


\subsection{Proof of Lemma \ref{lem1}}


Let $\{Y^{\len}_1,\ldots, Y^{\len}_\M\}$ 
be the set of $\M$ strings observed at the output of the channel.
In order to capture whether we are in the scenario of Figure~\ref{fig:points}(a) or (b), we let 
$T$ be the largest subset of $[1:\M] \defi \{1,\ldots,\M\}$ so that, for any $i,j \in T$, 
$
d_H \left( Y_i^{\len}, Y_j^{\len} \right) \geq \alpha \len,
$
where $d_H$ is the Hamming distance and 
$\alpha > 2p$. 
We assume that in case of ties, an arbitrary tie-breaking rule is used to define $T$ (the actual choice will not be relevant for the proof).

We will prove that, given the conditions in Lemma~\ref{lem1}, the following two bounds involving $\Ex|T|$ hold:
\al{
\text{(B1)\quad} & H\left(Y^{\ML}\right) \leq M + L \Ex|T|  \nonumber \\
& \quad \quad \quad  + (M-\Ex|T|)\left( \log \Ex|T| + \len H(\alpha) \right), \nonumber \\
\text{(B2)\quad} & H\left(S^M|X^{\ML},Y^{\ML}\right) \leq 
 M \log M \nonumber \\
&  \quad \quad \quad - \Ex|T| \log \Ex|T| + o(ML). \nonumber 
}
For large $\Ex|T|$, we are intuitively more often in the scenario in Figure~\ref{fig:points}(a), while Figure~\ref{fig:points}(b) corresponds to the case where $\Ex|T|$ is small.
The bounds above capture the tension between  $H\left(Y^{\ML}\right)$ and $H\left(S^M|X^{\ML},Y^{\ML}\right)$ because (B2) is decreasing in $\Ex|T|$, while (B1) is increasing in $\Ex|T|$ (as long as $\beta(1-H(\alpha)) \geq 1$).
Combining (B1) and (B2),
\al{
& H\left(Y^{\ML}\right) + H\left(S^M|X^{\ML},Y^{\ML}\right) \nonumber \\
& \quad \leq  \len \Ex|T| + (\M-\Ex|T|)\left( \log \Ex|T| + \len H(\alpha) \right) \nonumber \\
& \quad \quad \quad + \M \log \M - \Ex|T| \log \Ex|T| + o(\ML) \nonumber \\ 
& \quad = \Ex|T| L (1- H(\alpha))  + \M \log \Ex|T|- 2 \Ex|T| \log \Ex|T| \nonumber \\
& \quad \quad \quad  + \M \len H(\alpha) + \M \log \M + o(\ML).
\label{eq:boundET} 
}
Replacing $\Ex|T|$ with $x$ and ignoring the terms in this upper bound that do not involve $x$, we have the expression
\aln{
f(x) \defi x \beta(1-H(\alpha)) \log \M  + \M \log x - 2 x\log x.
}
Let $\gamma = \beta(1-H(\alpha))$.
For $x > 0$, we have 
\aln{
f'(x) & = \frac{1}{\ln(2)}\left( \gamma \log \M + \frac{M}{x} - 2\log x - 2 \right) \\
& > \frac{1}{\ln(2)}\left( \gamma \log \M - 2\log x - 2 \right) \\
& = \frac{2}{\ln(2)}\left( \log \frac{\M^{\gamma/2}}{x} - 1 \right).
}
Hence $f'(x) > 0$ if
\al{
x < \frac12 M^{\gamma/2}. \label{eq:maxx}
}
We see that, as long as $\gamma > 2$, the right-hand side of (\ref{eq:maxx}) is greater than $M$ for $M$ large enough.
This means that $f(x)$ is increasing for $1 \leq x \leq M$, and must attain its maximum at $f(M)$.
Notice that $x = \Ex|T| \leq M$.
Therefore, (\ref{eq:boundET}) can be upper-bounded by setting $x = \Ex|T| = M$, which yields
\aln{
H\left(Y^{\ML}\right) + H\left(S^M |X^{\ML},Y^{\ML}\right) \leq \ML + o(\ML).
}
Notice that this holds as long as
\aln{
\beta(1-H(\alpha)) > 2 \Leftrightarrow 1 - H(\alpha) - \frac{2}{\beta} > 0,
}
for some $\alpha > 2p$.
From the continuity of $H(\cdot)$, such $\alpha$ can be found if \eqref{eq:condition} is satisfied, proving the lemma.
Finally, we prove (B1) and (B2).

\vspace{2mm}

\noindent 
\emph{Proof of (B1):\;} 
Since $T$ is a deterministic function of $Y^{\ML}$ and can take at most $2^\M$ values, 
\al{
H\left(Y^{\ML}\right) & = H\left(Y^{\ML}, T \right) =  H\left( T \right) +  H\left(Y^{\ML} | T \right) \nonumber\\
& \leq M + \sum_{t \subseteq [1:\M]} \Pr\left( T = t \right) H\left(Y^{\ML} | T = t \right). \label{eq:bound3} 
}
Next we notice that, for a given $t$, we can write
\al{
H\left(Y^{\ML} | T = t \right) & \leq H\left( \{Y^{\len}_i : i \in t\} | T = t \right) \nonumber \\
& + H\left( \{Y^{\len}_i : i \not\in t\} | T = t, \{Y^{\len}_i : i \in t\} \right). \label{eq:bound4} 
}
The first term in  (\ref{eq:bound4}) is trivially bounded as 
\aln{
H\left( \{Y^{\len}_i : i \in t\} | T = t \right) \leq |t| L.
}
Each of the remaining length-$\len$ strings $Y^{\len}_i$ with $i \notin t$ must be within a distance $\alpha L$ from one of the strings in $\{Y^{\len}_i : i \in t\}$, from the definition of $T$.
Hence, conditioned on $\{Y^{\len}_i : i \in t\}$, each of them can only take at most $|t| |B(\alpha L)|$ values, where $B(\alpha L)$ is a Hamming ball of radius $\alpha \len$.
Since $|B(\alpha \len)| \leq 2^{\len H(\alpha)}$ for $\alpha < 0.5$, we bound the second term in (\ref{eq:bound4}) as
\aln{
& H\left( \{Y^{\len}_i : i \not\in t\} | T = t, \{Y^{\len}_i : i \in t\} \right) \\
& \quad \quad \quad \quad \leq (M-|t|) \left( \log |t| + \len H(\alpha) \right).
}
Using these bounds back in (\ref{eq:bound3}), we obtain
\al{
& H\left(Y^{\ML}\right)  \nonumber \\
& \quad \leq M + \Ex  \left[  L |T| + (M-|T|) \left( \log |T| + \len H(\alpha) \right) \right] \nonumber \\
& \quad \leq M + L \Ex|T| + (M-\Ex|T|)\left( \log \Ex|T| + \len H(\alpha) \right),
\label{eq:bound5}
}
where we used the fact that $(M-x) \log x$ is a concave function of $x$ and Jensen's inequality. \qed




\vspace{2mm}

\noindent \emph{Proof of (B2):\;}
Since $T$ is a deterministic function of $Y^{\ML}$,
\al{
& H\left(S^M|X^{\ML},Y^{\ML}\right) = H\left(S^M|X^{\ML},Y^{\ML}, T\right) \nonumber \\
& \quad  =  \sum_{t \subseteq [1:\M]} \Pr\left( T = t \right) H\left(S^M|X^{\ML},Y^{\ML}, T = t\right) \nonumber \\
& \quad  =  \sum_{t \subseteq [1:\M]} \Pr\left( T = t \right) \sum_{i}^M H\left(S(i)|X^{\ML},Y^{\ML}, T = t\right). \label{eq:bound6} 
}
Next we notice that the probability that $\delta \len$ or more errors occur in a single length-$\len$ string, for $\delta > p$, is at most $2^{-\len D(\delta || p)}$ by the Chernoff bound (where $D(\cdot || \cdot)$ is the binary KL divergence).
If we let $\E_i$ be the event that $d_H\left( X_{S(i)}^\len, Y_i^\len \right) \geq \delta L$, then we have
\aln{
\Pr(\E_i) \leq 2^{-\len D(\delta || p)} = \M^{-\beta D(\delta || p)}.
}
The conditional entropy term in (\ref{eq:bound6}) is upper bounded by
\al{
& H\left(S(i), \one_{\E_i}|X^{\ML},Y^{\ML}, T = t\right) \nonumber \\
& \quad  \leq H(\one_{\E_i}) + \Pr(\E_i | T = t) H\left(S(i)|X^{\ML},Y^{\ML}, T = t, \E_i \right) \nonumber \\
& \quad \quad \quad + \Pr(\bar\E_i | T = t) H\left(S(i)|X^{\ML},Y^{\ML}, T = t, \bar\E_i \right) \nonumber \\
& \quad \leq 1 + \Pr(\E_i | T = t) \log \M \nonumber \\
& \quad \quad \quad + H\left(S(i)|X^{\ML},Y^{\ML}, T = t, \bar\E_i \right)
\label{eq:bound7}
}
The final step is to bound the conditional entropy term in (\ref{eq:bound7}), for the case where $i \in t$.
Set $\delta = \alpha/2$. 
Conditioned on $\bar\E_i$, $d_H\left( X_{S(i)}^\len, Y_{i}^\len \right) < \alpha \len/2$.
Moreover, conditioned on $T=t$, for any $j \in t - \{i\}$, $d_H\left( Y_i^\len, Y_j^\len \right) \geq \alpha \len$.
Define the set 
%
%
%
%
%
%
%
%
\aln{
A_i = \{ j \st Y^{\len}_i \text{ is the closest output string in $t$ to }X^{\len}_j \}.
}
We claim that, if $i \in t$, $S(i)$ must be in $A_i$.
To see this notice that, for any $k \in t$, $k \ne i$, we have
\aln{
\alpha \len & \leq d_H \left( Y_i^{\len}, Y_k^{\len} \right) \\
& \leq d_H \left( X_{S(i)}^{\len}, Y_i^{\len} \right) + d_H \left( X_{S(i)}^{\len}, Y_k^{\len} \right)  \\
& < \alpha \len/2 + d_H \left( X_{S(i)}^{\len}, Y_k^{\len} \right),
}
implying that $d_H \left( X_{S(i)}^{\len}, Y_k^{\len} \right) > \alpha \len/2 \geq d_H \left( X_{S(i)}^{\len}, Y_i^{\len} \right)$.
%
Therefore, $S(i)$ for each output string $Y^{\len}_i$ with $i \in t$, can take at most $|A_i|$ values.
Hence we have
\al{
& \sum_{i}^M H\left(S(i)|X^{\ML},Y^{\ML}, T = t, \bar\E_i \right) \nonumber \\
& \quad \quad \quad \leq \sum_{i \not\in t}H(S(i)) + \sum_{i \in t} \log |A_i| \nonumber \\
& \quad \quad \quad \leq (M-|t|) \log M + \sum_{i \in t} \log |A_i| \nonumber \\
& \quad \quad \quad \leq (M-|t|) \log M +|t| \log (M/|t|) \nonumber \\
& \quad \quad \quad = M \log M - |t| \log |t|, \label{eq:boundAi}
}
where the last inequality follows because $\sum_{i \in t} |A_i| = M$, and the sum is maximized by $|A_i| = \M/|t|$.
Combining (\ref{eq:bound6}), (\ref{eq:bound7}), and (\ref{eq:boundAi}), we obtain
\al{
& H\left(S^M|X^{\ML},Y^{\ML}\right) =  \nonumber \\
& \quad  =  \sum_{i}^M \sum_{t \subseteq [1:\M]} \Pr\left( T = t \right)\left[ 1 + \Pr(\E_i | T = t) \log \M  \right] \nonumber \\
& \quad  \;\; +  \sum_{t \subseteq [1:\M]} \Pr\left( T = t \right) \sum_{i}^M H\left(S(i)|X^{\ML},Y^{\ML}, T = t, \bar \E_i \right) \nonumber \\
& \quad  =  M + \log M\sum_{i=1}^M \Pr(\E_i) + M \log M - \Ex\left[|T| \log |T|\right] \nonumber \\
& \quad  \leq  M + M \log M (1+M^{-\beta D(\delta || p)} ) - \Ex|T| \log \Ex|T| \nonumber
}
%
%
%
where we used Jensen's inequality in the last step.
Since $ \M^{-\beta D(\delta || p)} \to 0$ as $\M \to \infty$,
$\M^{-\beta D(\delta || p)} \M \log \M = o(\ML)$, concluding the proof. \qed

\section{Discussion and Extensions} \label{sec:discussion}

While the parameter regime of $(p,\beta)$ in the blue region of Figure~\ref{fig:capacity} is practically the most relevant one, an interesting question for future work 
 is whether expression~(\ref{eq:capacity}) is still the capacity of the BSC-shuffling channel if $\beta$ and $p$ do not satisfy (\ref{eq:condition}) (i.e., the gray region in Figure~\ref{fig:capacity}).
Notice that this is a high-noise, short-block regime, and it is reasonable to imagine that coding across the different sequences can be helpful and an index-based approach might not be optimal. 


\vspace{-0.2cm}
\subsection{General symmetric channels}
\vspace{-0.1cm}

If we view the capacity expression in (\ref{eq:capacity}) as $\Cbsc -1/\beta$, it is natural to wonder whether for a different noisy channel with capacity $C_{\text{noisy}}$, the correspoinding noisy shuffling channel would have capacity $C_{\text{noisy}}-1/\beta$.
As it turns out, the converse proof in Section~\ref{sec:converse} can be extended to the class of \emph{symmetric} discrete memoryless channels (as described in \cite[ch. 7.2]{CoverThomas}). 

Consider a noisy shuffling channel formed by a symmetric discrete memoryless channel (SDMC) with output alphabet $\mathcal{Y}$, followed by a shuffling channel.
We focus on the same asymptotic regime from Section~\ref{sec:problem}.
Then we have:

\begin{theorem}
\label{thm:extension}
If $\beta$ is large enough, 
the capacity of the SDMC-shuffling channel is given by
\al{
C = C_{\text{SDMC}} - 1/\beta. \label{eq:capacity2}
}
Moreover, if $\beta \leq \log|\mathcal{Y}|$, $C = 0$.
\end{theorem}

For symmetric channels, capacity is achieved by making the distribution of the output uniform, which allows an analogous result to Lemma~\ref{lem1} to be obtained. 
Notice that how large $\beta$ needs to be depends on the specific channel transition matrix.

\vspace{-0.1cm}
\section{Concluding Remarks}
\vspace{-0.1cm}

In this paper we took steps towards the understanding of the fundamental limits of DNA-based storage systems. 
We proposed a simple model capturing the fact that molecules are stored in an unordered fashion, are short, and are corrupted by individual base errors. 
Our results show that a simple index-based coding scheme is asymptotically optimal.

While the model captures (moderate) substitution errors which are the prevalent error source on a nucleotide level of current DNA storage systems, the current generation of systems relies on \emph{low-error} synthesis and sequencing technologies that are relatively expensive and limited in speed.
A key idea towards developing the next-generation of DNA storage systems is to employ \emph{high-error}, but cheaper and faster synthesis and sequencing technologies such as light-directed maskless synthesis of DNA and nanopore sequencing. 
Such systems induce a significant amount of insertion and deletion errors. Thus, and important area of further investigation is to understand the capacity of channels which introduce deletions and insertions as well.

In addition, in practice, one can introduce physical redundancy (i.e., duplicate molecules) in storage and redundancy in the DNA sequencing. Thus, an interesting question is to understand how to efficiently code for channels where we have several unordered, independently perturbed copies of each sequence. 


\end{document}